\documentclass{article}
\usepackage{amsmath,amssymb,amsfonts,amsthm}
\usepackage[russian]{babel}
\usepackage{color}
\usepackage{graphicx}
\usepackage{wrapfig}

\begin{document}

\flushbottom

\renewcommand{\figurename}{Fig.}
\def\refname{References}
\def\proofname{Proof}

\newtheorem{teo}{Theorem}
\newtheorem{pro}{Proposition}
\newtheorem{rem}{Remark}

\def\tens#1{\ensuremath{\mathsf{#1}}}

\if@mathematic
   \def\vec#1{\ensuremath{\mathchoice
                     {\mbox{\boldmath$\displaystyle\mathbf{#1}$}}
                     {\mbox{\boldmath$\textstyle\mathbf{#1}$}}
                     {\mbox{\boldmath$\scriptstyle\mathbf{#1}$}}
                     {\mbox{\boldmath$\scriptscriptstyle\mathbf{#1}$}}}}
\else
   \def\vec#1{\ensuremath{\mathchoice
                     {\mbox{\boldmath$\displaystyle\mathbf{#1}$}}
                     {\mbox{\boldmath$\textstyle\mathbf{#1}$}}
                     {\mbox{\boldmath$\scriptstyle\mathbf{#1}$}}
                     {\mbox{\boldmath$\scriptscriptstyle\mathbf{#1}$}}}}
\fi

\newcommand*{\greysquare}{\textcolor{gray}{\blacksquare}}
\newcommand{\bs}{\boldsymbol}
\newcommand{\diag}{\mathop{\rm diag}\nolimits}
\newcommand{\wt}{\widetilde}
\newcommand{\q}{\quad}
\newcommand{\const}{\rm const}
\newcommand{\vfi}{\varphi}
\newcommand{\bM}{\bs M}
\newcommand{\bL}{\bs L}
\newcommand{\bN}{\bs N}
\newcommand{\mR}{\mathbb{R}}
\newcommand{\eps}{\varepsilon}

\begin{center}
{\Large\bf Generalizations of the Kovalevskaya case and~quaternions\\}

\bigskip


{\large\bf Ivan~A.\,Bizyaev$^1$,
Alexey~V.\,Borisov$^2$,
Ivan~S.\,Mamaev$^3$\\}
\end{center}

\begin{quote}
\begin{small}
\noindent
$^{1,2,3}$ Steklov Mathematical Institute, Russian Academy of Sciences,\\
ul.~Gubkina~8, Moscow, 119991 Russia\\[2mm]
$^1$ E-mail: bizaev\_90@mail.ru\\
$^2$ E-mail: borisov@rcd.ru\\
$^3$ E-mail: mamaev@rcd.ru
\end{small}

\bigskip
\bigskip

\begin{small}
\textbf{Abstract.} This paper provides a detailed description of various
reduction schemes in rigid body dynamics. Analysis of one of such
nontrivial\linebreak reductions makes it possible to order the cases
already found and to obtain new generalizations of the Kovalevskaya case
to $e(3)$. We note that the above reduction allows one to obtain in a
natural way some singular additive terms which were proposed earlier by
D.\,N.\,Goryachev.

\smallskip

\textbf{Keywords} rigid body dynamics, quaternions, reduction, cyclic
coordinates, Kovalevskaya top

\smallskip

\end{small}
\end{quote}

\clearpage				
\section{Introduction}	
Two possible integrable generalizations of the classical Kovalevskaya
top\footnote{There is an extensive literature on the classical
Kovalevskaya top (see, e.g., \cite{DTT, Bobenko})} are well known from
rigid body dynamics. One of them involves the introduction of additional
terms to the system with two degrees of freedom on the algebra $e(3)$, for
example, a gyrostatic parameter or terms added to the potential, which do
not break the symmetry of the field relative to the fixed axis. These
additive terms were examined in detail by Yehia~\cite{Yehia, Yehia02},
Valent~\cite{valent} and others.

The other type of generalization involves the introduction of additional
fields (along with the gravitational field), for example, a magnetic and
homogeneous electric field. In the general case, the fields are assumed to
be transversal to each other. In contrast to the gravitational field,
additional fields do not allow a usual reduction by the precession angle
(which is cyclic due to the invariance of the system under rotations about
a fixed axis), in this case it is necessary to investigate a system with
three degrees of freedom. A general integrable case for this system was
obtained by A.\,G.\,Reyman and M.\,A.\,Semenov-Tian-Shansky~\cite{resem},
and a generalization of this case was obtained by A.\,V.\,Tsiganov and
V.\,V.\,Sokolov~\cite{ST}. This system possesses a quadratic integral and
a fourth-degree integral (for bifurcation analysis of this case see the
work of\linebreak M.\,P.\,Kharlamov \cite{Kharlamov2014}). In some cases
presented in \cite{ya, BM}, the quadratic integral reduces to a linear
one, and a constructive order reduction is possible (see also~\cite{BM}).

An interesting fact was that for a constructive reduction it is convenient
to use quaternions, which were introduced and advocated by W.\,Hamilton to
solve various mechanics problems. In the paper by Borisov and
Mamaev~\cite{BM} (1997), an explicit process of reduction using
quaternions for the equations of rigid body dynamics was described and
isomorphisms between different integrable systems were revealed. However,
these results remained little-known (apparently because they were
published only in the Russian language), and, as a consequence,
publications still appear in which the connections between systems, as
described in~\cite{BM}, are ignored. For example, the author
of~\cite{Tsiganov} presents a ``new'' integrable system which, as it turns
out, can be obtained using reduction from a general integrable system as
obtained earlier in~\cite{ST}. Moreover, the author of~\cite{Kharlamov}
presents a separation of variables which can be obtained using the above
isomorphism~\cite{BM} and from the separation found earlier in~\cite{Ra} .

The goal of this paper is to give, once again, a more detailed description
of various reduction schemes in rigid body dynamics, which are of interest
in themselves and are presented only in the book~\cite{DTT}, which has
also been published only in the Russian language. Analysis of one
nontrivial reduction makes it possible to order the cases already found
and to obtain new\linebreak generalizations of the Kovalevskaya case to
$e(3)$. We note that the above\linebreak reduction allows one to obtain
for this case in a natural way some singular additive terms which were
proposed earlier in the work of D.\,N.\,Goryachev~\cite{97, 98}. To
conclude, we note that the quaternion equations presented in~\cite{BM}
and~\cite{DTT} are still poorly understood, although they can be used for
various algebraic and geometric methods of integration, for example,
in~\cite{gen} they were used to construct an ${\bf L}$-${\bf A}$ pair of
the Goryachev\,--\,Chaplygin top.

\section{Equations of motion}

Consider a rigid body rotating in a potential force field about a fixed
point~$O$. The configuration space, which is a set of all positions of the
rigid body, is the Lie group~$SO(3)$, and we can take, for example,
\glossary{Эйлер} the {\itshape Euler
angles}~$\theta,\varphi,\psi$~\cite{DTT} as (local) coordinates specifying
the position of the rigid body.

\begin{figure}[!ht]
\centering
\includegraphics[totalheight=4.5cm]{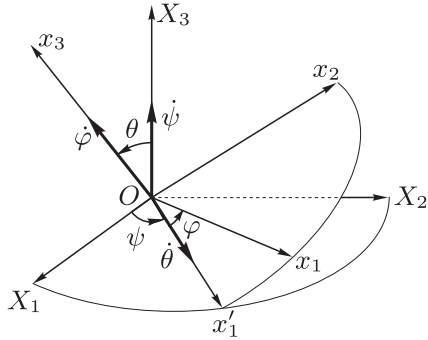}
\caption{Euler angles.}
\label{fig3}
\end{figure}

To specify them, we define two coordinate systems with origin at the fixed
point $O$:
\begin{itemize}
\item[{--}] a fixed coordinate system $OX_1X_2X_3$,
\item[{--}] a moving coordinate system $Ox_1x_2x_3$ rigidly attached
    to the rotating rigid body (Fig.~\ref{fig3}).
\end{itemize}

The transformation from the fixed axes to the moving axes is given by the
orthogonal matrix ${\bf Q} \in SO(3)$ (matrix of direction cosines), which
is defined by three successive rotations:
$$
\begin{gathered}
{\bf Q}_\theta=
\begin{pmatrix}
1 & 0 & 0 \\
0 & \cos\theta & \!-\!\sin\theta \\
0 & \sin\theta & \cos\theta
\end{pmatrix}\!\!, \quad
{\bf Q}_\varphi=
\begin{pmatrix}
\cos\varphi & -\sin\varphi & 0 \\
\sin\varphi & \cos\varphi & 0 \\
0 & 0 & 1
\end{pmatrix}\!\!, \\
{\bf Q}_\psi=
\begin{pmatrix}
\cos\psi & -\sin\psi & 0 \\
\sin\psi & \cos\psi & 0 \\
0 & 0 & 1
\end{pmatrix}\!\!,
\end{gathered}
$$
\begin{align}
{\bf Q}&={\bf Q}_\psi {\bf Q}_\theta {\bf Q}_\varphi={}\notag\\
&{}=
\text{\small
\arraycolsep=2pt
$\left(\hskip-\arraycolsep\begin{array}{ccc}
\cos\varphi\cos\psi{-}\cos\theta\sin\psi\sin\varphi &
-\sin\varphi\cos\psi{-}\cos\theta\sin\psi\cos\varphi  &   \sin\theta\sin\psi \\
\cos\varphi\sin\psi{+}\cos\theta\cos\psi\sin\varphi  & -\sin\varphi\sin\psi
{+}\cos\theta\cos\psi\cos\varphi   & -\sin\theta\cos\psi \\
\sin\varphi\sin\theta   & \cos\varphi\sin\theta  & \cos\theta \\
\end{array}\hskip-\arraycolsep\right)$}={} \notag\\
\label{eq2.3}
&{}=\begin{pmatrix}
\alpha_1 & \alpha_2 & \alpha_3 \\
\beta_1 & \beta_2 & \beta_3 \\
\gamma_1 & \gamma_2 & \gamma_3
\end{pmatrix}\!\!.
\end{align}

The rows of this matrix, $\bs \alpha=(\alpha_1, \alpha_2, \alpha_3)$, $\bs \beta=(\beta_1, \beta_2,
\beta_3)$, $\bs \gamma=(\gamma_1, \gamma_2, \gamma_3)$, are the unit vectors of the fixed axes $OX_1X_2X_3$ projected onto the moving axes
$Ox_1x_2x_3$. Since they have a clear geometric meaning, they are used in what follows to elucidate the physical meaning of forces acting on the body.

Supplementing these position variables with the corresponding canonical momenta $p_\theta$, $p_\varphi$, $p_\psi$, we write the equations of motion
of the body in the canonical Hamiltonian form
$$
\begin{gathered}
\dot \theta =\frac{\partial H}{\partial p_\theta}, \quad \dot \varphi=\frac{\partial H}{\partial p_\varphi}, \quad
\dot \psi=\frac{\partial H}{\partial p_\psi},\\
\dot p_\theta=-\frac{\partial H}{\partial \theta}, \quad \dot p_\varphi=-\frac{\partial H}{\partial \theta}, \quad
\dot p_\psi=-\frac{\partial H}{\partial \psi}.
\end{gathered}
$$

As a rule, it is not convenient to use equations in this form to search for and analyze integrable caseв, so we transform them to an appropriate form.
To do this, we shall use as configuration variables the quaternions $\lambda=(\lambda_0, \lambda_1, \lambda_2, \lambda_3)$
with the unit norm
$$
\lambda_0^2+\lambda_1^2+\lambda_2^2+\lambda_3^2=1,
$$
which are also called the Rodrigues\,--\,Hamilton parameters. Their relation with the Euler angles is given by

\begin{equation}
\label{Q2-5}
\begin{aligned}
\lambda_{0}&=\cos\frac{\theta}{2}\cos\frac{\psi+\varphi}{2},\quad &
\lambda_{1}&=\sin\frac{\theta}{2}\cos\frac{\psi-\varphi}{2},\\
\lambda_{2}&=\sin\frac{\theta}{2}\sin\frac{\psi-\varphi}{2},\quad &
\lambda_{3}&=\cos\frac{\theta}{2}\sin\frac{\psi+\varphi}{2}.
\end{aligned}
\end{equation}
Multiplication in the group $SO(3)$ is consistent with multiplication of the quaternions, and the matrix of direction cosines is written as
\begin{equation*}
\label{}
{\bf Q} =
\arraycolsep=4pt
\text{\small$\left(\hskip-\arraycolsep
\begin{array}{ccc}
\lambda_0^2+\lambda_1^2-\lambda_2^2-\lambda_3^2&
2(\lambda_1\lambda_2-\lambda_0\lambda_3) &
2(\lambda_0\lambda_2+\lambda_1\lambda_3) \\[3pt]
2(\lambda_0\lambda_3+\lambda_1\lambda_2) &
\lambda_0^2-\lambda_1^2+\lambda_2^2-\lambda_3^2&
2(\lambda_2\lambda_3-\lambda_0\lambda_1) \\[3pt]
2(\lambda_1\lambda_3-\lambda_0\lambda_2)  &
2(\lambda_0\lambda_1+\lambda_2\lambda_3)  &
\lambda_0^2-\lambda_1^2-\lambda_2^2+\lambda_3^2
\end{array}\hskip-\arraycolsep\right)$}.
\end{equation*}

\begin{rem}
From the geometrical point of view, quaternions with unit norm form a three-dimensional sphere $S^3$, which doubly covers the group $SO(3)$.
It is this fact that relations~\eqref{Q2-5} express.
\end{rem}

It is more convenient to use, instead of canonical momenta, the projections of angular momentum $\bs M$ onto the moving axes $Ox_1x_2x_3$, which are defined as
$$
\begin{gathered}
M_1=\frac{\sin\varphi}{\sin\theta}(p_\psi-p_\varphi\cos\theta)+p_\theta\cos\varphi,\quad
M_2=\frac{\cos\varphi}{\sin\theta}(p_\psi-p_\varphi\cos\theta)-p_\theta\sin\varphi, \quad \\
M_3=p_\varphi.
\end{gathered}
$$
In this case, the kinetic energy of the rigid body is a quadratic form with constant coefficients
$$
T=\frac{1}{2}(\bs M, {\bf A}\bs M), \quad {\bf A}={\bf I}^{-1},
$$
where $\bf I$~is the tensor of inertia of the body relative to the point $O$ in the moving coordinate system $Ox_1x_2x_3$; if $Ox_1, x_2, x_3$~are the principal
axes of
inertia, then ${\bf I}=\diag(I_1, I_2, I_3)$ and hence ${\bf A}=\diag (a_1, a_2,
a_3)$, $a_i=I_i^{-1}$.

In the new variables the Poisson structure turns out to be linear (Lie\,--\,Poisson bracket)

\begin{equation}
\label{Q2-7}
\begin{gathered}
\{M_{i},M_{j}\}=-\varepsilon_{ijk}M_{k}, \quad \{\lambda_0, \lambda_i\}=\{\lambda_i, \lambda_j\}=0, \\
\{M_{i},\lambda_{0}\}=\frac12\lambda_{i}, \quad
\{M_{i},\lambda_{j}\}=-\frac12(\varepsilon_{ijk}\lambda_{k}+\delta_{ij}
\lambda_{0}). \qquad
\end{gathered}
i, j, k=1, 2, 3
\end{equation}
It turns out to be degenerate and possesses the unique Casimir function
$$
C_0(\lambda)=\lambda_0^2+\lambda_1^2+\lambda_2^2+\lambda_3^2.
$$

The equations of motion of the body in the new variables can be represented in vector
form:
\begin{equation}
\label{eq02}
\begin{gathered}
\dot{\bs{M}}=\bs{M} \times \frac{\partial H}{\partial \bs{M}} + \frac{1}{2}\bs{\lambda}\times \frac{\partial H }{\partial \bs{\lambda}}
 + \frac{1}{2}\bs{\lambda}\times \frac{\partial H }{\partial\lambda_0} - \frac{1}{2}\lambda_0\frac{\partial H}{\partial \bs{\lambda}}, \\
 \dot{\lambda}_0=-\frac{1}{2}\left( \bs{\lambda}, \frac{\partial H}{\partial \bs{M}} \right), \quad
 \dot{\bs{\lambda}}=\frac{1}{2}\bs{\lambda}\times\frac{\partial H}{\partial \bs{M}} + \frac{1}{2}\lambda_0 \frac{\partial H}{\partial \bs{M}},
\end{gathered}
\end{equation}
where $\bs{\lambda}=(\lambda_1, \lambda_2, \lambda_3)$.
In what follows we consider Hamiltonians of the form
\begin{equation}
\label{eq2.2}
H=\frac{1}{2}(\bs M, {\bf A}\bs M)+\big(\bs M, \bs W(\lambda)\big)+U(\lambda),
\end{equation}
where $U(\lambda)$ and $\bs W(\lambda)$~are, respectively, the scalar and vector potentials describing the interaction of the body with the external fields.

\begin{rem}
There is a connection between the Rodrigues\,--\,Hamilton parameters and the direction cosines $\bs{\alpha}$,
$\bs{\beta}$, $\bs{\gamma}$:
\begin{equation}
\label{eq14}
\begin{gathered}
\lambda_0^2=\frac{1 + \alpha_1 + \beta_2 + \gamma_3}{4}, \quad
\lambda_1^2=\frac{1 + \alpha_1 - \beta_2 - \gamma_3}{4}, \\
\lambda_2^2=\frac{1 - \alpha_1 + \beta_2 - \gamma_3}{4}, \quad
\lambda_3^2=\frac{1 - \alpha_1 - \beta_2 + \gamma_3}{4}.
\end{gathered}
\end{equation}
\end{rem}

\section{Reduction}

We now consider three cases in which the system~\eqref{eq02} is invariant under the (Hamiltonian) action of the rotation group $S^1$.
The corresponding Hamiltonians generating these actions have the form
$$
H_1=p_\psi, \quad H_2=p_\varphi, \quad H_3=p_\psi-p_\varphi.
$$

In the matrix representation~\eqref{eq2.3} the action on $SO(3)$, which is generated by the Hamiltonian $H_1$, is given by multiplication on the left by the
rotation matrix:
$$
g_t({\bf Q})={\bf S}_t{\bf Q}, \quad
{\bf S}_t=
\begin{pmatrix}
\cos t & -\sin t & 0 \\
\sin t & \cos t & 0 \\
0 & 0 & 1
\end{pmatrix}\!\!.
$$

For the Hamiltonian $H_2$ this action is a multiplication on the left:
$$
g_t({\bf Q})={\bf Q}{\bf S}_t
$$

For the Hamiltonian $H_3$ the corresponding action is defined by the relation
$$
g_t({\bf Q})={\bf S}_t{\bf Q}{\bf S}_{-t}.
$$

As is well known, a reduction (of order) of the system is possible in this case. Moreover, in all three
cases the reduced system can be represented in a natural way in Hamiltonian
form on the zero orbit of the coalgebra $e^*(3)$. In other words, one can choose the variables of the reduced system $\bs L=(L_1, L_2, L_3)$, $\bs n=(n_1, n_2,
n_3)$ in such a way that they form a Lie\,--\,Poisson bracket of the form
\begin{equation}
\label{eq2.4}
\{L_i, L_j\}=-\varepsilon_{ijk}L_k, \quad \{L_i, n_j\}=-\varepsilon_{ijk}n_k, \quad \{n_i, n_j\}=0,
\end{equation}
and the level set of the Casimir functions is fixed as follows:
\begin{equation}
\label{eq2.5}
C_1=(\bs n, \bs n)=1, \quad C_2=(\bs L, \bs n)=0.
\end{equation}

The equations of motion in the new variables have the well-known form
\begin{equation}
\label{eq2.6}
\dot{\bs{L}}=\bs{L}\times\frac{\partial H}{\partial \bs{L}} +
\bs{n}\times\frac{\partial H}{\partial \bs{n}}, \quad
\dot{\bs{n}}=\bs{n}\times\frac{\partial H}{\partial \bs{L}}.
\end{equation}

This will allow us to establish a relation between different integrable cases.

\subsection{The area integral $H_1=p_\psi$}

Symmetries leading to such an integral are natural: they are due to the
invariance of the external field under rotations about some fixed axis.
Such axisymmetric fields include homogeneous fields, in particular, a
gravitational field. The\linebreak precession angle~$\psi$ is a cyclic
variable.

Let us choose the invariants of this action as follows:
\begin{gather*}
n_1=2(\lambda_1\lambda_2-\lambda_0\lambda_2)=\sin\theta \sin\varphi, \quad
n_2=2(\lambda_0\lambda_1+\lambda_2\lambda_3)=\sin\theta \cos\varphi,\\
n_3=\lambda_0^2-\lambda_1^2-\lambda_2^2+\lambda_3^2=\cos\theta,\\
L_1=M_1-\frac{(\bs M, \bs n)n_1}{n_1^2+n_2^2}, \quad
L_2=M_2-\frac{(\bs M, \bs n)n_2}{n_1^2+n_2^2}, \quad
L_3=M_3,
\end{gather*}
where $\bs n=(n_1, n_2, n_3)=\bs\gamma$~is the last column of the matrix
$\bf Q$. This column has a simple physical meaning: it is a unit vector
directed along the symmetry axis in the coordinate system $Ox_1x_2x_3$. In
this case, the area integral is represented as
$$
H_1=p_\psi=(\bs M, \bs n).
$$

A straightforward verification shows that the variables $\bs L$, $\bs
n$ satisfy\linebreak relations~\eqref{eq2.4} and~\eqref{eq2.5}.

Using the relation $\{H, H_1\}=0$, we find conditions under which the system with the Hamiltonian~\eqref{eq02} admits
this symmetry group. Expressed in terms of Euler angles, these conditions have the simplest form:
$$
\frac{\partial W_i}{\partial \psi}=0, \quad \frac{\partial U}{\partial \psi}=0, \quad i=1, 2, 3.
$$

Thus, at a fixed value of the area integral $p_\psi=c$
we obtain a Hamiltonian of the reduced system in the form
$$
\begin{gathered}
H=\frac{1}{2}(\bs L, {\bf A}\bs L)+(\bs L, \bs W_r)+U_r,\\
\bs W_r=\bs W+c {\bf A}\bs \tau, \quad
U_r=U+c(\bs r, \bs W)+\frac{c^2}{2}(\bs \tau, {\bf A}\bs \tau),\\
\bs \tau=\Bigg(\frac{n_1}{n_1^2+n_2^2},\, \frac{n_2}{n_1^2+n_2^2}, \, 0\Bigg).
\end{gathered}
$$
As can be seen, when $c \ne 0$, this system has singularities on the
Poisson sphere $\bs n^2=1$ at the points $\bs n_+=(0, 0, 1)$ and $\bs
n_-=(0, 0, -1)$.

\subsection{The Lagrange integral $H_2=p_\varphi$}

This integral is a projection of the angular momentum on the body-fixed
axis $Ox_3$:
$$
H_2=M_3.
$$
In this case, the rigid body must be dynamically symmetric:
$$
I_1=I_2,
$$
where without loss of generality we set $I_1=I_2=1$. Moreover, it is necessary that the force field also be invariant under rotation about the axis
of dynamical symmetry, and so in the general form the Hamiltonian of the system can be represented as
\begin{equation}
\label{eq3.3}
H\!\!=\!\!\frac{1}{2}(M_1^2+M_2^2+a_3M_3^2)\!+\!\wt{W}_1(\theta, \psi)(M_1\gamma_1+M_2\gamma_2+M_3\gamma_3)\!+\!\wt{W}_3(\theta, \psi)M_3+
U(\theta, \psi).
\end{equation}

As the invariants of the action of the symmetry group we have to choose
the following variables:
$$
\begin{gathered}
n_1=\alpha_3=\sin\theta\sin\psi, \quad n_2=\beta_3=-\sin\theta\cos\psi, \quad
n_3=\gamma_3=\cos\theta;\\
L_1=M_3\frac{\alpha_3}{\alpha_3^2+\beta_3^2}-(\bs M, \bs \alpha), \quad
L_2=M_3\frac{\beta_3}{\alpha_3^2+\beta_3^2}-(\bs M, \bs \beta), \quad L_3=-(\bs M, \bs \gamma).
\end{gathered}
$$
As above, these variables satisfy
relations~\eqref{eq2.4} and~\eqref{eq2.5}.

On the fixed level set of the Lagrange integral $M_3=c$, from~\eqref{eq3.3} we obtain a Hamiltonian
of the reduced system in the form
$$
H=\frac{1}{2}\bs L^2-L_3\left(\wt{W}_1(\bs n)-c\frac{n_3}{n_1^2+n_2^2}\right)+\frac{c^2}{2}\frac{n_3^2}{n_1^2+n_2^2}+U(\bs n),
$$
where the insignificant constants have been omitted.

\subsection{The integral $H_3=p_\psi-p_\varphi$}

In quaternion variables this integral is represented as
$$
H_3=2M_3(\lambda_1^2 + \lambda_2^2) - 2(M_1\lambda_1 + M_2\lambda_2)\lambda_3 +2 (M_1\lambda_2 - M_2\lambda_1)\lambda_0.
$$
\begin{rem}
We note that in the direction cosines this integral has the form
$$
H_3=M_3 - (\bs{M}, \bs{\gamma}).
$$ 	
\end{rem}	

First of all, we find out under what conditions a system with
the\linebreak Hamiltonian~\eqref{eq2.2} admits this integral. From the
relation $\{H, H_3\}=0$ we find that the most general form of the
Hamiltonian is
\begin{equation}
\label{eq06}
\begin{aligned}
H&=\frac{1}{2}(M_1^2 + M_2^2 + a_3M_3^2) + \wt{W}_1(\lambda_0, \lambda_3)(M_1\lambda_1 + M_2\lambda_2) -{}\\
&{}- \wt{W}_2(\lambda_0, \lambda_3) (M_1\lambda_2 - M_2\lambda_1) - \wt{W}_3(\lambda_0, \lambda_3)M_3 + U(\lambda_0, \lambda_3),
\end{aligned}
\end{equation}
where $a$ is an arbitrary constant and $\wt{W}_i(\lambda_0, \lambda_3)$, $i=1,2,3$ and  $U(\lambda_0, \lambda_3)$ are the functions characterizing
the vector and scalar potentials of the external field. The cyclic variable is $\psi - \varphi$.

This implies, in particular, that for the existence of this symmetry
group it is necessary that the body have dynamical symmetry: $I_1=I_2$. (Without loss of generality, in~\eqref{eq06} we have set
$I_1=1$, $I_3=a_3^{-1}$).

The invariants of the action generated by the Hamiltonian $H_2$ can be chosen as
follows:
\begin{equation}
\begin{gathered}
\label{eq3.1}
n_1\!\!=\!\!\lambda_3=\cos\frac{\theta}{2}\sin\frac{\psi+\varphi}{2}, \, n_2\!\!=\!\!\lambda_0=\cos\frac{\theta}{2}\cos\frac{\psi+\varphi}
{2}, \,
n_3\!\!=\!\!\sqrt{\lambda_1^2+\lambda_2^2}=\sin\frac{\theta}{2}, \\
L_1=\frac{2}{\sqrt{\lambda_1^2+\lambda_2^2}}\left(\frac{H_3}{2C_0}\lambda_3-M_1\lambda_1-M_2\lambda_2\right), \\
\quad L_2=
\frac{2}{\sqrt{\lambda_1^2+\lambda_2^2}}\left(\frac{H_3}{2C_0}\lambda_0+M_1\lambda_2-M_2\lambda_1\right)
\end{gathered}
\end{equation}
$$
L_3=\frac{H_3}{C_0}+2M_3.
$$

For the variables \eqref{eq3.1} relations~\eqref{eq2.4},~\eqref{eq2.5} and~\eqref{eq2.6} hold, i.e., as in the previous case the reduced system
is represented only on the orbit of the coalgebra $e^*(3)$ on the zero level set of the area integral.

On the fixed level set of the first integrals $H_3=c$ and $C_0=1$, in the new variables the Hamiltonian~\eqref{eq06} can be rewritten, up to insignificant
constants, as follows:
\begin{equation}
\label{eq05}
\begin{gathered}
H=\frac{1}{8}(L_1^2 + L_2^2 + a_3L_3^2)+\frac{1}{2}(\bs L, \bs{\wt{W}_r})+U_r, \\
\bs{\wt{W}_r}=\left(n_1\wt{W}_1-\frac{c}{2n_3}n_1, \quad n_2\wt{W}_2-\frac{c}{2n_3}n_2, \quad \wt{W}_3-\frac{c}{2}a_3\right) \\
U_r=U(n_1, n_2)-\frac{c}{2}(n_1\wt{W}_1+n_2\wt{W}_2+\wt{W}_3)+\frac{c^2}{8 n_3^2},
\end{gathered}
\end{equation}
where $\bs {\wt W}=(\wt{W}_1, \wt{W}_2, \wt{W}_3).$ As we can see, the reduced system has singularities on the equator $n_3=0$ of the Poisson sphere $\bs n^2=1$.

We refer the reader to the recent and extensive work \cite{BMred}, which
is devoted to general methods of reduction in nonholonomic systems
(describing, for example, the rolling motion of rigid bodies). It would be
interesting to investigate\linebreak a~nonholonomic (non-Hamiltonian)
analog of the reduction described in the above-mentioned work. Such an
analog arises, for example, in the analysis of the rolling motion of a
rigid body in the presence of homogeneous force fields. Such problems have
not yet been considered in nonholonomic mechanics.

\section{The Kovalevskaya case}

Consider a particular case of the Hamiltonian \eqref{eq06} that
corresponds to the Kovalevskaya top lying in a potential field and that in
the direction cosines $\bs{\alpha}, \bs{\beta}, \bs{\gamma}$ has the form
\begin{equation}
	\begin{gathered}
		\label{b01}
		H=\frac{1}{2}(M_1^2 + M_2^2 + 2M_3^2) + c_1M_3 + c_2(M_1\alpha_3 + M_2\beta_3) - \\- c_2(\alpha_1 + \beta_2)M_3 + c_3(\alpha_2 - \beta_1).
	\end{gathered}
\end{equation}
In this case, the equations of motion possess, along with $H_2$, an
additional integral of degree 4 in momenta, which can be obtained from the
${\bf L}$-${\bf A}$ pair presented by V.\,V.\,Sokolov and A.\,V.\,Tsiganov
\cite{ST} and generalizing the earlier ${\bf L}$-${\bf A}$ pair of
A.\,G.\,Reyman, M.\,A.\,Semenov-Tian-Shansky~\cite{resem}.
\begin{rem}
We note that, generally speaking, the authors of \cite{ST} presented
a~more general case of the Kovalevskaya top in which the equations of
motion possess an additional integral of degree 2 and 4 in momenta (the
explicit form of the integral is presented in \cite{Ra2}).
\end{rem}

After passing to the quaternions we find
$$
\wt{W}_1=2c_2\lambda_3, \quad \wt{W}_2=-2c_2\lambda_0, \quad \wt{W}_3= 2(\lambda_0^2 - \lambda_3^2)c_2 - c_1, \quad
U=-4c_3\lambda_0\lambda_3.
$$
Further, by applying the reduction procedure described in the previous
section, we obtain an (integrable) system on $e(3)$ on the zero level set
of the area integral. In order to compare the resulting system with those
obtained earlier, we make the canonical change of variables
$$
\begin{gathered}
L_1\to\frac{L_1 + L_2}{\sqrt{2}}, \quad L_2\to\frac{L_1 - L_2}{\sqrt{2}}, \q L_3\to L_3,\\
n_1\to\frac{n_1 + n_2}{\sqrt{2}}, \quad n_2\to\frac{n_1 - n_2}{\sqrt{2}}, \q n_3\to n_3.
\end{gathered}
$$
As a result, the Hamiltonian can be represented as
\begin{equation}
\label{eq09}
\begin{aligned}
H&=\frac{1}{8}(L_1^2 + L_2^2 + 2L_3^2) - \frac{1}{4}(c + 2c_1)L_3 + {}\\
&{}+ c_2(n_2n_3L_1 + n_1n_3L_2 - 2 n_1n_2L_3 ) -
2c_3(n_1^2 - n_2^2)+ \frac{c^2}{8n_3^2}.
\end{aligned}
\end{equation}
In this case, the additional integral has the form
$$
\begin{gathered}
F=\left( L_1^2 - L_2^2 + 8c_2n_3(L_1n_2 - L_2n_1) - 16c_3n_3^2 +
\frac{c^2(n_1^2 + n_2^2)}{n_3^2} \right)^2+{}\\
{}+4\left( L_1L_2 - \frac{c^2n_1n_2}{n_3^2}\right)^2 +
4(2c_1 + c)(L_3 - 2c_1 - c )\left(L_1^2 + L_2^2 +
c^2\left(1 + \frac{1}{n_3^2}\right)\!\!\right)+{}\\
{}+ 64c_3(2c_1+c)(L_1n_1 - L_2n_2) +{}\\
{}+ 32c_2(2c_1+c)\left(L_1^2n_1n_2 + L_1L_2n_3^2 + L_2^2n_1n_2 + c^2\frac{n_1n_2}{n_3^2}\right)\!.
\end{gathered}
$$
Various particular cases of the Hamiltonian \eqref{eq09} were presented earlier.
\begin{itemize}
	\item[---] If the relations
	$$
	c_2=0, \quad c=-2c_1
	$$	
	hold, then the Hamiltonian \eqref{eq09} reduces to the Goryachev case, for which a separation of variables was performed in
\cite{Ra}.
    A separation of variables in the initial system (i.e., with the Hamiltonian \eqref{b01}), also under the above condition,
	but without the above analogy, was performed in \cite{Kharlamov}.
	It is obvious that in this case the separation in \cite{Kharlamov}
	coincides with that in~\cite{Ra}.
	\item[---] The last section of \cite{Kharlamov} discusses the case
	$$
	c=0, \quad c_1=0, \quad c_2=0.
	$$
	It follows from the above isomorphism that it is equivalent to the Chaplygin case integrated by Chaplygin
himself. This isomorphism was found for the first time in \cite{BM}.
	\item[---] Under the conditions
	$$
	c_1=0, \quad c=0
	$$
	the additional integral was found earlier by A.V. Tsiganov in \cite{Tsiganov}. As can be seen, this integral is not new and can be obtained from
the system \cite{ST} using the reduction procedure described above.
\end{itemize}	

\begin{rem}
There are a lot of publications by H.M.Yehia (and his colleagues) on the
generalization of the Kovalevskaya case. A weak point of these
publications is the absence of Hamiltonian formalism of the problem.
Because of this many generalizations presented by Yehia are dependent and
can be obtained by the simplest algebraic transformation. This aspect of
his work is discussed in the book \cite{DTT} and in the recent paper
\cite{Bir}. 	
\end{rem}

\section{Quaternion Euler\,--\,Poisson equations}

Let us consider the case of equations of motion of a rigid body with a
potential that is linear not in the direction cosines, but in the
quaternions
\begin{equation}
\label{kvat-eq-1}
H=\frac{1}{2}(\bs M, {\bf A}\bs M)+\sum_{i=0}^3r_i\lambda_i,\quad {\bf A}={\rm diag}(a_1,a_2,a_3), \quad r_i=\const,
\end{equation}
assuming that the equations of motion have the form \eqref{eq02}. We note
that such potentials are not encountered in mechanics, since their
dependence on the position of the body is ambiguous \eqref{eq14}. Problems
of quantum mechanics, dynamics of point masses in curved
space~$S^3$~\cite{BorisovMamaev}, as well as some formal methods for
constructing {\bf L}-{\bf A}-pairs~\cite{BorisovMamaev} can be regarded as
a motivation for considering such equations. Moreover, it turns out that
the order reduction of the system~\eqref{kvat-eq-1} leads to the standard
Euler\,--\,Poisson equations with~additional terms having different
physical interpretations.

An interesting singularity of the system~\eqref{kvat-eq-1} is that using
transformations linear in~$\lambda_i$, the general form of the potential
\begin{equation}
\label{kvat-eq-2}
V=\sum_{i=0}^3 r_i\lambda_i
\end{equation}
can be reduced to the form
\begin{equation}
\label{kvat-eq-3}
V=r_0\lambda_0.
\end{equation}
Indeed, linear transformations of the quaternion space~$\lambda_i$ (which
do not change the commutator relations and the norm of the quaternion) of
the form
\begin{equation}
\label{lin-kvater}
\begin{aligned}
\wt\lambda_0&=R^{-1}(r_0\lambda_0+r_1\lambda_1+r_2\lambda_2+r_3\lambda_3),\\
\wt\lambda_1&=R^{-1}(r_0\lambda_1-r_1\lambda_0-r_2\lambda_3+r_3\lambda_2),\\
\wt\lambda_2&=R^{-1}(r_0\lambda_2+r_1\lambda_3-r_2\lambda_0-r_3\lambda_1),\\
\wt\lambda_3&=R^{-1}(r_0\lambda_3-r_1\lambda_2+r_2\lambda_1-r_3\lambda_0),\\
R^2\span = r_0^2+r_1^2+r_2^2+r_3^2
\end{aligned}
\end{equation}
reduce the potential~\eqref{kvat-eq-2} to the form~\eqref{kvat-eq-3}. The
existence of such a linear transformation is a remarkable singularity of
the quaternion variables and of the bracket~\eqref{Q2-7}; it has no
analogs for the brackets of the algebra~$e(3)$ and~$so(4)$.

In the dynamically asymmetric case~$a_1\ne a_2\ne a_3\ne a_1$ the
system~\eqref{kvat-eq-1} is apparently nonintegrable and neither of the
two necessary additional integrals exists. It would be interesting to use
the Kovalevskaya method and other methods for finding additional first
integrals to investigate \eqref{kvat-eq-1}.

For $a_1=a_2$ there always exists the linear integral
\begin{equation}
\label{kvat-eq-4}
\begin{aligned}
H_3&=M_3(r_0^2+r_1^2+r_2^2+r_3^2)+N_3(r_1^2+r_2^2-r_0^2-r_3^2)+\\
{}&\quad + 2N_2(r_1r_0-r_3r_2)-2N_1(r_1r_2-r_0r_3),
\end{aligned}
\end{equation}
where~$N_i$ are the projections of angular momentum onto the fixed axes
$$
N_1=(\bs{M}, \bs{\alpha}), \quad N_2=(\bs{M}, \bs{\beta}), \quad N_1=(\bs{M}, \bs{\gamma}).
$$
Under the conditions~$r_1=r_2=r_3=0$ this integral takes the natural form
\begin{equation}
\label{kvat-eq-5}
H_3=M_3-N_3
\end{equation}
and corresponds to the cyclic
variable~$\vfi+\psi$. The reduction described above leads to a Hamiltonian system on the algebra~$e(3)$ with zero value of the area
integral~$(\bM,\bs\gamma)=0$ and with the Hamiltonian
\begin{equation}
\label{kvat-eq-6}
H=\frac12(M_1^2+M_2^2+a_3M_3^2)+c(a_3-1)M_3+r_0\gamma_2+
\frac12\frac{c^2}{\gamma_3^2}.
\end{equation}

We present here integrable cases of the system~\eqref{kvat-eq-1} which turn out to be equivalent to the integrable cases
of the system~\eqref{kvat-eq-6}.

\paragraph[The spherical top $(a_1=a_2=a_3)$]
{The spherical top $(a_1=a_2=a_3)$.}\index{Волчок!шаровой}
The Hamiltonian has the form$$
H=\frac12 \bM^2+r_0\lambda_0,
$$
and, as shown in~\cite{BorisovMamaev}, the system is equivalent to the
problem of the motion of a~material point over a three-dimensional
sphere~$S^3$. Since the potential depends only on~$\lambda_0$, it can be
assumed that the material point moves in the field of a~fixed center
placed in the northern (southern) pole, and the force of interaction
depends only on the distance to it (analog of the problem in the central
field for~$\mR^3)$. As in the planar case, the angular momentum vector of
the particle is preserved:
\begin{equation}
\label{kvat-eq-7}
\bL=\frac12\,(\bN-\bM)=\const,
\end{equation}
where~$\bN$ is the angular momentum vector in the fixed axes.

The components of the vector~$\bL$ form the algebra~$so(3)$: $\{L_i,L_j\}=
\eps_{ijk}L_k$, and~{\em the integrability is noncommutative}; moreover,
the system possesses\linebreak a redundant set of integrals and its
three-dimensional tori are foliated by two-dimensional tori.

\paragraph[The Kovalevskaya case]
{The Kovalevskaya case.}
The Hamiltonian and the additional integral involutive to~$F_1$ (of degree 4) have the form
\begin{align}
H&=\frac12(M_1^2+M_2^2+2M_3^2)+r_0\lambda_0,\notag\\
\label{kvat-eq-8}
F&=(M_1N_1+M_2N_2+2r_0\lambda_0 )^2+(N_1M_2-N_2M_1-2r_0\lambda_3)^2+\\
{}+&(N_3-M_3)\bigl(M_3(\bM^2-M_3N_3)+
2r_0(M_2\lambda_1-M_1\lambda_2+\frac{\lambda_0}{2}(M_3-N_3))\bigr).\notag
\end{align}

\paragraph[The Goryachev\,--\,Chaplygin case]{The Goryachev\,--\,Chaplygin case.}
The Hamiltonian and the additional integral have the form
\begin{equation}
\label{kvat-eq-9}
\begin{aligned}
H&=\frac12(M_1^2+M_2^2+4M_3^2)+r_0\lambda_0,\\
F&=M_3(M_1^2+M_2^2)+r_0(M_2\lambda_1-M_1\lambda_2).
\end{aligned}
\end{equation}

Somewhat unexpected is the circumstance that the Lagrange and Hess cases
cannot be generalized to the system~\eqref{kvat-eq-1}. We note that for
the quaternion Euler\,--\,Poisson equations both the Kovalevskaya case and
the Goryachev\,--\,Chaplygin case are general integrable cases.

\section{Generalization of the quaternion cases\\
of Kovalevskaya and Goryachev\,--\,Chaplygin}

Let us consider a generalization of the quaternion case of Kovalevskaya
\begin{equation}
\label{eq10}
H=\frac{1}{2}(M_1^2 + M_2^2 + 2M_3^2) + c_1M_3 + 2c_2\lambda_0\lambda_3 + r_0\lambda_0 .
\end{equation}
After reduction (by eliminating the cyclic variable $\varphi+\psi$) the
Hamiltonian \eqref{eq10} can be represented as
\begin{equation}
\label{eq11}
H=\frac{1}{8}(L_1^2 + L_2^2 + 2L_3^2) - \frac{1}{4}(c + 2c_1)L_3 + r_0n_2   + 2c_2n_1n_2 + \frac{c^2}{8n_3^2}.
\end{equation}
The integrability of the Hamiltonian \eqref{eq11} on the zero level set of
the area integral was shown by H.\,M.\,Yehia \cite{Yehia}. In this case,
the additional integral has the form
$$
\begin{gathered}
F_3=\left( L_1^2 - L_2^2 - c^2\frac{n_1^2 - n_2^2}{n_3^2} + 8r_0n_2\right) ^2 +\\+
4\left( L_1L_2 - c^2\frac{n_1n_2}{n_3^2} - 4r_0n_1  + 4c_2n_3^2\right) ^2 + \\+
4(c+2c_1)(L_3 - c - 2c_1)\left( L_1^2 + L_2^2 + c^2\left( 1 + \frac{1}{n_3^2}\right) \right)  - \\- 16(c+2c_1)n_3(c_2n_2L_1 + L_2(r_0+c_2n_1)).
\end{gathered}
$$
We present the second generalization of the Kovalevskaya case
\begin{equation}
\label{eq15}
H=\frac{1}{2}(M_1^2 + M_2^2 + 2M_3^2) +c_1(M_1\lambda_1 + M_2\lambda_2 + M_3\lambda_3) +  c_2M_3  + r_0\lambda_0.
\end{equation}

After reduction on the zero level set of the area integral of the algebra
$e(3)$ the Hamiltonian \eqref{eq15} can be represented as
\begin{equation}
\label{eq16}
H=\frac{1}{8}(L_1^2 + L_2^2 + 2L_3^2) + \frac{1}{4}(c + 2c_2)L_3 - \frac{1}{2}c_1(n_1L_3 - n_3L_1) + \frac{c^2}{8n_3^2} + r_0n_2.
\end{equation}
The family \eqref{eq16} was presented by H.\,M.\,Yehia \cite{Yehia02}, and
the additional integral in this case has the form
$$
\begin{gathered}
F=\left(L_1^2 - L_2^2 + 4c_1(L_1n_3 - L_3n_1) + 4c_1^2 + 8r_0n_2 - \frac{c^2(n_1^2 - n_2^2)}{n_3^2} \right)^2 + \\+
4\left( L_1L_2 + 2c_1(L_2n_3 - L_3n_2) - 4r_0n_1 - \frac{c^2n_1n_2}{n_3^2} \right)^2 + \\  + 4(c+2c_2)(L_3 -c -2c_2)(L_1^2+L_2^2) +
16(c+2c_2)(c_1n_2L_1L_2 - c_1n_1L_2^2 - \\
-c_1^2L_3 - 2r_0n_3L_2) -\frac{4 c^2(c+2c_2)}{n_3^2}( 4c_1n_1 + c+2c_2  - (1+n_3^2)L_3 ).
\end{gathered}
$$
To conclude, we consider a generalization of the quaternion case of
Goryachev\,--\,Chaplygin
\begin{equation}
\label{eq17}
H=\frac{1}{2}(M_1^2 + M_2^2 + 4 M_3^2) + c_1(M_1\lambda_1 + M_2\lambda_2 + 2M_3\lambda_3) + c_2 M_3 + r_0\lambda_0.
\end{equation}
After reduction the Hamiltonian has the form
$$
H\!\!=\!\!\frac{1}{8}(L_1^2 + L_2^2 + 4 L_3^2 ) - \frac{1}{4}(2c_2 + 3c)L_3 + \frac{1}{2}c_1(2n_1L_3 - n_3L_1) + \frac{c^2}{8n_3^2} - \frac{1}{2}cc_1n_1 + r_0n_2.
$$
The additional integral in this case was found in \cite{ST}.\goodbreak

In this paper we have shown interrelations between different
integrable\linebreak systems resulting from reduction. These
interrelations can be used in the inverse problem, that of recovering the
dynamics. This procedure allows one to construct from an integrable system
with two degrees of freedom an integrable three-degree-of-freedom system
possessing additional symmetry. (In principle this procedure can also be
applied to nonintegrable systems.) This issue is partially discussed in
the book~\cite{DTT}. The issue of integrable generalizations of
the\linebreak Kovalevskaya case in which the potential is a superposition
of terms linear and quadratic (linear in the direction cosines) in
quaternions remains open.

\bigskip

The authors express their gratitude to P.\,E.\,Ryabov and Yu.\,N.\,Fedorov
for useful discussions.

\bigskip

This research was supported by the Russian Scientific Foundation (project
No 14-50-00005) at the Steklov Mathematical Institute of the Russian
Academy of Sciences.

\end{document}